# Some results of analysis of source position time series

Zinovy Malkin

27 January 2009

**Abstract.** Source position time series produced by International VLBI Service for Geodesy and astrometry (IVS) Analysis Centers were analyzed. These series was computed using different software and analysis strategy. Comparison of this series showed that they have considerably different scatter and systematic behavior. Based on the inspection of all the series, new sources were identified as sources with irregular (non-random) position variations. Two statistics used to estimate the noise level in the time series, namely RMS and ADEV were compared.

## 1. Introduction

This memo is intended to contribute to two ICRF-2 preparation steps in accordance with the ICRF-2 Working Plan:
- ranking of sources by time series statistics,
- compiling list of specially handling sources.

In accordance with the ICRF-2 Working Plan, in this memo, only quantitative source stability indices are considered. Since source position variations are sometimes pretty complicated, a visual inspection still may be needed to make a final decision on the quality of a specific source (Gordon et al. 2008).

In this study we considered 33 source position time series submitted in 2007–2009 by IVS analysis centers AUS, BKG, DGF, GSF, IAA, MAO, OPA, SAI, SHA and USN. Some of initially submitted series were replaced by extended versions, some series don not contain session ID which makes them not suitable for some kind of analysis, some series contain too few sources, etc. Finally, 15 series were selected for analysis. These 15 series present various software and analysis strategies which makes them interesting and useful for comparison.

The following preliminary operations were made:
- only position with at least 5 observations were selected,
- only time series with at least 5 positions (epochs) were selected,
- for velocity analysis, only series with time span at least 5 years were selected,
- error floor of 20 µas for source position uncertainty was applied; which practically does not affect most of the series except the IAA ones.

## 2. Preliminary analysis

To better understand the result of time series analysis we tried to perform a comparison of methods used in various Analysis Centers for computation of the time series. Preliminary result is presented in Table 1 which still should be checked and completed.

Table 1. The main characteristics of source position time series.

| Series software | Time span | EOP | Stations | Sources | Comment |
|---|---|---|---|---|---|
| bkg000c Calc/Solve | 1984.0 2007.5 | all | global | part global, part local | |
| dgf000f Calc/Solve | 1984.0 2007.6 | all | no | local with NNR and 3-param transformation (?) | |
| dgf000g Occam LS | 1984.0 2007.6 | all | no | local with NNR and 4-param transformation (?) | |
| gsf001a Calc/Solve | 1979.6 2007.9 | ? | ? | part global, part local | |
| gsf003a Calc/Solve | 1979.6 2008.7 | ? | ? | part global, part local | mobile and small networks excluded |
| iaa001b QUASAR | 1979.6 2008.6 | all | ? | single session solutions for every source | |
| iaa001c QUASAR | 1979.6 2008.6 | no | no | single session solutions for every source | |
| mao000b SteelBreeze | 1980.3 2007.3 | ? | ? | local (?) | |
| mao006a SteelBreeze | 1979.6 2008.7 | ? | ? | local (?) | |
| opa000b Calc/Solve | 1984.0 2008.0 | all | no | session solutions | |
| opa002a Calc/Solve | 1984.0 2008.7 | LOD, nutation | local | part global, part local | |
| sai000b ARIADNA | 1984.0 2008.0 | ? | ? | session solutions with NNR for selected sources | |
| sha006a Calc/Solve | 1979.6 2008.8 | ? | ? | part local. part global | |
| usn000d Calc/Solve | 1979.6 2007.4 | LOD, nutation | local | part local, part global | |
| usn001a Calc/Solve | 1979.6 2007.4 | no | no | session solutions | |

To get a preliminary impression about differences of the position time series produced by the participated Analysis Centers, source position time series were visually inspected. As was already pointed out in Malkin (2008b), different time series show different source position scatter and, for some sources, different long-term behavior. Two examples are given in Figs. 1 and 2. Three series mao000b, opa000b and usn001a show clear change point in DE around 2003.0 for both sources. Very similar break at the same epoch can be seen also for 0851+202, 0003–0660, 0727–115 and some other sources. Somewhat other situation can be seen for 1741–038 (Fig. 2) where two DE series opa000b and usn001a show a supplement change point around 1993.0 or maybe long-period signal. Commonly speaking, e.g. usn001a series shows strong irregular position variations for many sources, which are not seen in usn000d or most of other series.

The epoch of jump around 2003.0 may be connected with at least two events: the earthquake near Gilmore Creek and the start of regular operations at Svetloe. The first event seems to be the most important and worth careful investigation. Macmillan & Cohen (2004) and Titov & Tregoning (2005) detected an effect of the Alaska 2002 earthquake on EOP estimates. Evidently, there will be also an impact of different treatment of GILCREEK position on source coordinates. As a test, a source position time series w/o GILCREEK may be computed.

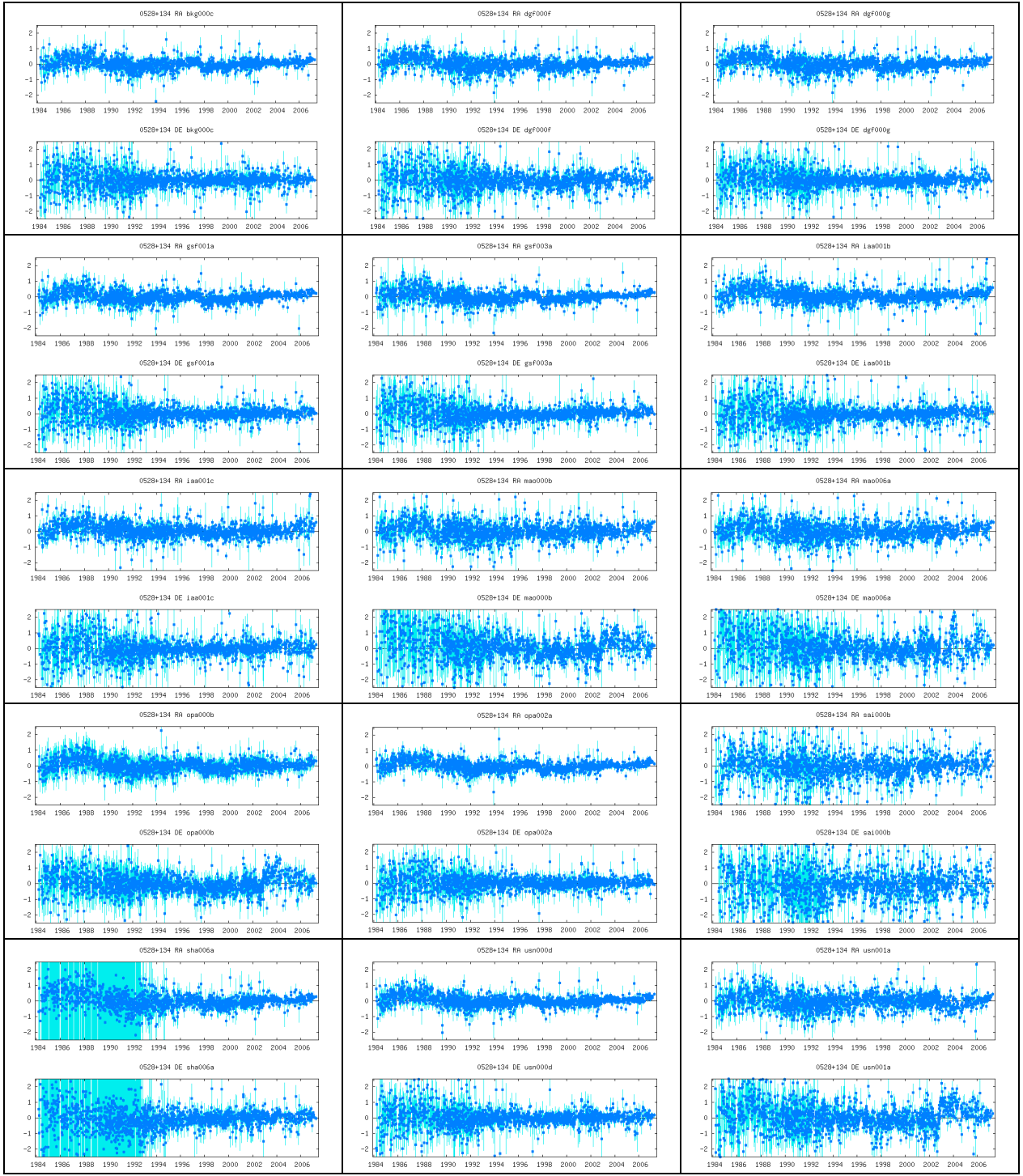

Fig. 1.— Position time series for 0528+134. It should be mentioned that all the newest time series extending till 2008 (gsf, iaa, mao, opa, sha) show a clear change point, especially in RA, at the epoch around 2007.0.

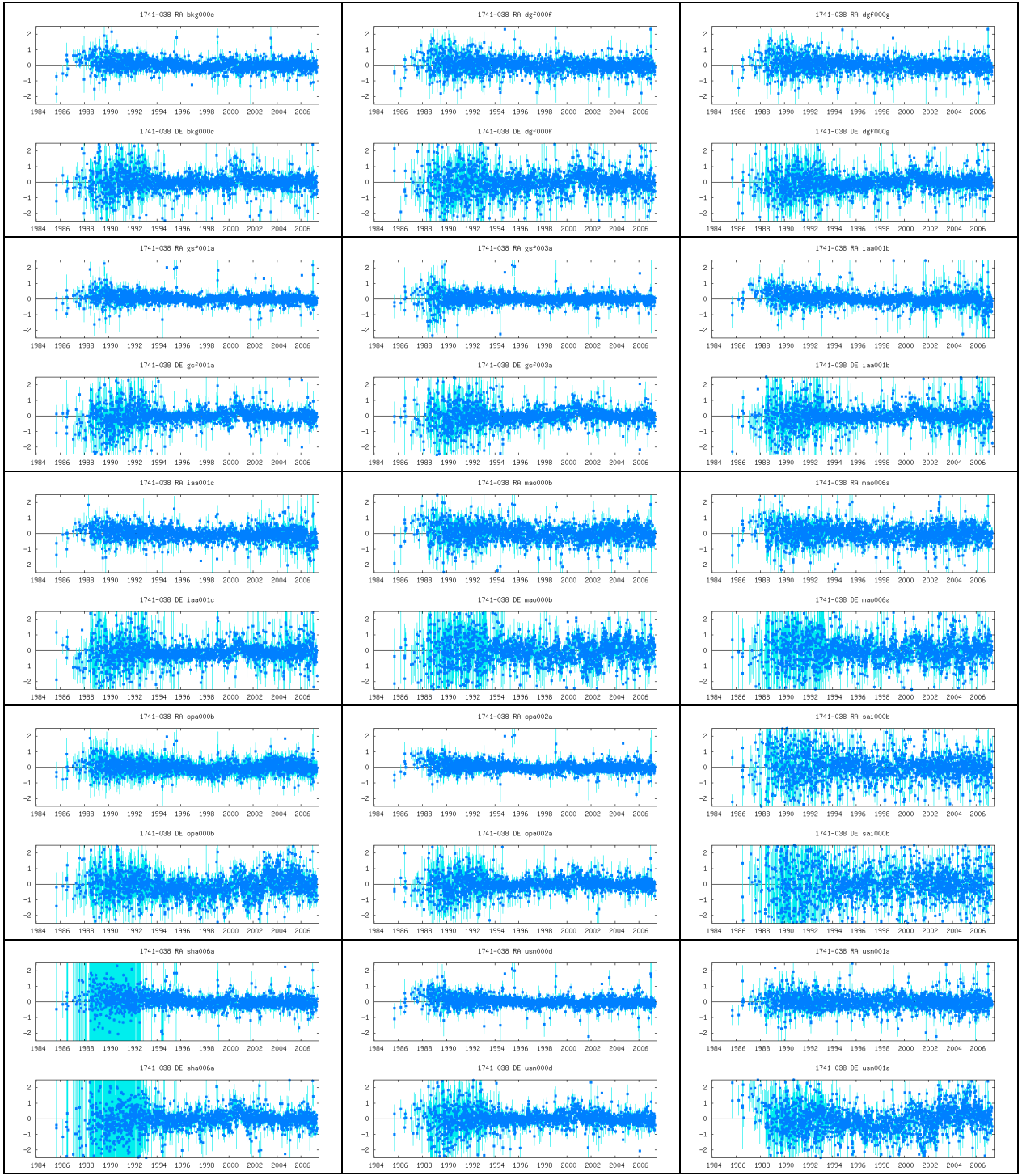

Fig. 2. Position time series for 1741–038.

There is no room in this memo to analyze all the time series. However, a conclusion can be made that the methods used for generation of source position time series considerably influence not only the scatter but also systematic behavior of source position variations.
Both examples show also different noise level for different series, which will be quantified in the next section.

# 3. Ranking of sources by instability indices

From several scatter or instability indices used for analysis of source position time series we used two ones weighted root-mean-square residuals of the session source position w.r.t. weighted mean position, where weights are inversely proportional to the square of formal error (WRMS) and weighted Allan deviation (WADEV) computed as proposed in Malkin (2008a). The combined RA and DE (2D) results are used for source ranking.

Both used statistics, WRMS and WADEV have own advantages and disadvantages. WRMS estimate includes contribution of systematic (e.g. trend-like) and low-frequency (quasi-) periodic variations. WADEV is not affected by slow position variations; however, it may give inadequate estimate of source position stability in a case of jumps in the time series and small position variation at time intervals between jumps. To get more general index of source position variations, composite index computed as the sum of WRMS and WADEV was used for final source ranking.

The computations were performed in two steps. At the first step, source positions for common epochs were selected from 15 series listed above. Several statistics were computed for these "common-epochs" series, two of them are presented in Table 2. The table allows us to compare the noise of each series, and later use this data for series selection at the next iteration of this work. The last row of Table 1 shows statistics for the weighted average series.

Average series was computed and used for final source ranking. Indeed, it should be considered as the first iteration only since, as shown above, systematic difference between series may exist depending on time series generation method, and these differences should be explained and eliminated. Besides, a set of time series used for averaging should be accurately selected. However, as our tests showed, this does not much influence the source ranking results, which may be more sensitive to session and time span and interval selection, cf. e.g. Tables 3 and 4 which present the results of source ranking based on WRMS+WADEV statistics of average time series computed for common only and for all epochs.

Table 2. 2D WRMS/WADEV for the most frequently observed sources and median values for all the sources, µas. N – number of common epochs (sessions).

| Series | 0552+398 N=2168 | 0923+392 N=1919 | 1741-038 N=1868 | 0851+202 N=1678 | 0727-115 N=1675 | 0528+134 N=1653 | 1749+096 N=1516 | Median |
|---|---|---|---|---|---|---|---|---|
| bkg000c | 207 / 209 | 316 / 251 | 385 / 420 | 324 / 330 | 413 / 472 | 362 / 370 | 312 / 353 | 457 / 515 |
| dgf000f | 235 / 233 | 344 / 289 | 547 / 546 | 381 / 372 | 558 / 585 | 453 / 436 | 433 / 433 | 512 / 557 |
| dgf000g | 180 / 186 | 302 / 248 | 437 / 450 | 296 / 306 | 476 / 524 | 368 / 368 | 323 / 350 | 442 / 499 |
| gsf001a | 161 / 161 | 258 / 208 | 294 / 316 | 275 / 262 | 365 / 410 | 334 / 337 | 234 / 261 | 394 / 449 |
| gsf003a | 163 / 166 | 254 / 203 | 311 / 334 | 273 / 273 | 350 / 397 | 321 / 321 | 238 / 261 | 400 / 456 |
| iaa001b | 200 / 210 | 317 / 245 | 413 / 426 | 380 / 336 | 442 / 495 | 397 / 409 | 303 / 348 | 464 / 529 |
| iaa001c | 284 / 288 | 340 / 289 | 486 / 473 | 386 / 384 | 502 / 552 | 425 / 444 | 393 / 407 | 522 / 585 |
| mao000b | 442 / 434 | 511 / 484 | 728 / 713 | 612 / 566 | 705 / 738 | 682 / 633 | 657 / 620 | 675 / 718 |
| mao006a | 486 / 463 | 563 / 519 | 931 / 803 | 775 / 636 | 937 / 889 | 868 / 706 | 857 / 709 | 740 / 767 |
| opa000b | 355 / 338 | 435 / 393 | 742 / 665 | 519 / 473 | 669 / 671 | 606 / 526 | 650 / 563 | 577 / 588 |
| opa002a | 174 / 178 | 278 / 218 | 307 / 330 | 284 / 270 | 344 / 394 | 328 / 330 | 242 / 281 | 382 / 430 |
| sai000b | 504 / 581 | 624 / 718 | 968 /1113 | 699 / 766 | 1146/1372 | 852 / 978 | 789 / 915 | 906 /1076 |
| sha006a | 185 / 188 | 249 / 217 | 354 /387 | 278 / 278 | 468 / 505 | 389 / 366 | 296 / 337 | 446 / 501 |
| usn000d | 167 / 172 | 279 / 219 | 331 / 354 | 289 / 281 | 379 / 432 | 338 / 346 | 260 / 295 | 402 / 468 |
| usn001a | 445 / 429 | 522 / 472 | 722 / 711 | 620 / 550 | 670 / 693 | 690 / 630 | 660 / 620 | 680 / 732 |
| Average | 159 / 162 | 285 / 206 | 337 / 345 | 284 / 284 | 363 / 391 | 351 / 340 | 301 / 280 | 413 /455 |

Table 3.  Ranking of sources by WRMS+WADEV, common sessions, 411 sources.

| Source | Value | Source | Value | Source | Value | Source | Value | Source | Value | Source | Value |
|---|---|---|---|---|---|---|---|---|---|---|---|
| 1357+769 | 0.2814 | 1435+638 | 0.5604 | 0336-019 | 0.7306 | 0148+274 | 0.9025 | 0201+113 | 1.1625 | 0111+021 | 1.5634 |
| 1300+580 | 0.2824 | 2136+141 | 0.5623 | 0836+710 | 0.7316 | 1228+126 | 0.9050 | 2252-090 | 1.1698 | 0457+024 | 1.5646 |
| 0059+581 | 0.2964 | 2229+695 | 0.5638 | 1954+513 | 0.7351 | 0221+067 | 0.9059 | 1555+001 | 1.1721 | 2128-123 | 1.5691 |
| 0133+476 | 0.3124 | 1849+670 | 0.5660 | 1633+382 | 0.7379 | 1404+286 | 0.9210 | 1023+131 | 1.1743 | 1133+704 | 1.5734 |
| 0602+673 | 0.3128 | 1717+178 | 0.5662 | 1144+402 | 0.7444 | 0048-097 | 0.9237 | 1548+056 | 1.1766 | 0839+187 | 1.5779 |
| 0552+398 | 0.3205 | 0851+202 | 0.5679 | 0716+714 | 0.7487 | 1123+264 | 0.9243 | 0636+680 | 1.1778 | 0402-362 | 1.5836 |
| 1807+698 | 0.3272 | 1111+149 | 0.5707 | 1637+826 | 0.7528 | 2235+731 | 0.9253 | 2008-159 | 1.1818 | 1302-102 | 1.6166 |
| 0749+540 | 0.3301 | 0828+493 | 0.5811 | 0736+017 | 0.7541 | 1705+135 | 0.9264 | 2335-027 | 1.1866 | 0237-027 | 1.6287 |
| 1448+762 | 0.3378 | 0234+285 | 0.5891 | 0727-115 | 0.7547 | 2121+053 | 0.9269 | 0539-057 | 1.1941 | 0814+425 | 1.6419 |
| 1739+522 | 0.3511 | 0008-264 | 0.5917 | 0637-752 | 0.7588 | 0610+260 | 0.9295 | 1215+303 | 1.1956 | 2345-167 | 1.6554 |
| 1751+288 | 0.3581 | 2223-052 | 0.5940 | 0754+100 | 0.7590 | 1958-179 | 0.9300 | 0013-005 | 1.1993 | 1806-458 | 1.6636 |
| 0642+449 | 0.3645 | 2113+293 | 0.5946 | 2253+417 | 0.7609 | 1413+135 | 0.9318 | 2155-152 | 1.2107 | 2059+034 | 1.6974 |
| 1039+811 | 0.3735 | 1749+701 | 0.5947 | 2021+317 | 0.7613 | 0130-171 | 0.9357 | 2021+614 | 1.2109 | 0738+313 | 1.7130 |
| 1800+440 | 0.3849 | 1342+662 | 0.5963 | 1928+738 | 0.7614 | 0502+049 | 0.9374 | 0409+229 | 1.2206 | 0723-008 | 1.7365 |
| 0955+476 | 0.3882 | 0722+145 | 0.6024 | 0805+046 | 0.7623 | 0454-234 | 0.9459 | 2227-088 | 1.2303 | 1402-012 | 1.7392 |
| 1803+784 | 0.3942 | 1758+388 | 0.6043 | 0536+145 | 0.7702 | 1929+226 | 0.9530 | 1104-445 | 1.2335 | 0238-084 | 1.7556 |
| 0224+671 | 0.3957 | 1315+346 | 0.6073 | 0600+177 | 0.7713 | 1830+285 | 0.9567 | 1255-316 | 1.2390 | 2254+024 | 1.7749 |
| 0035+413 | 0.3977 | 0615+820 | 0.6115 | 0556+238 | 0.7778 | 1226+373 | 0.9681 | 0406+121 | 1.2409 | 1451-400 | 1.7804 |
| 1417+385 | 0.3992 | 1508+572 | 0.6117 | 1124-186 | 0.7790 | 0823+033 | 0.9682 | 1614+051 | 1.2460 | 0524-460 | 1.7839 |
| 1044+719 | 0.4053 | 1600+335 | 0.6124 | 1652+398 | 0.7820 | 2144+092 | 0.9700 | 2243-123 | 1.2470 | 0405-123 | 1.8240 |
| 1639+230 | 0.4057 | 1705+456 | 0.6185 | 1504+377 | 0.7824 | 1546+027 | 0.9714 | 2344+092 | 1.2601 | 1253-055 | 1.8596 |
| 1655+077 | 0.4066 | 0820+560 | 0.6185 | 0123+257 | 0.7870 | 1920-211 | 0.9742 | 1522+155 | 1.2607 | 1313-333 | 1.8607 |
| 1611+343 | 0.4131 | 0748+126 | 0.6190 | 2254-074 | 0.7879 | 0426+273 | 0.9746 | 1705+018 | 1.2674 | 1349-439 | 1.8738 |
| 1638+398 | 0.4144 | 1923+210 | 0.6205 | 0601+245 | 0.7900 | 0003-066 | 0.9748 | 1213-172 | 1.2713 | 0256-075 | 1.9108 |
| 0110+495 | 0.4202 | 2201+315 | 0.6221 | 0322+222 | 0.7907 | 1743+173 | 0.9784 | 2230+114 | 1.2850 | 0422-380 | 1.9134 |
| 1213+350 | 0.4300 | 1459+480 | 0.6311 | 0003+380 | 0.7948 | 1532+016 | 0.9843 | 0420-014 | 1.3026 | 0430+052 | 1.9177 |
| 1011+250 | 0.4316 | 1732+389 | 0.6312 | 1342+663 | 0.7950 | 0951+693 | 0.9852 | 1004+141 | 1.3050 | 1932+204 | 1.9218 |
| 2037+511 | 0.4357 | 1538+149 | 0.6375 | 2351+456 | 0.7966 | 2052-474 | 0.9933 | 2251+158 | 1.3053 | 1040+123 | 1.9279 |
| 0804+499 | 0.4372 | 0954+658 | 0.6404 | 1746+470 | 0.7991 | 1751+441 | 0.9937 | 1354+195 | 1.3063 | 2255-282 | 1.9439 |
| 0149+218 | 0.4390 | 0945+408 | 0.6406 | 0215+015 | 0.7994 | 0138-097 | 1.0011 | 0239+108 | 1.3094 | 2106+143 | 1.9479 |
| 1221+809 | 0.4470 | 0821+621 | 0.6409 | 0151+474 | 0.8025 | 0537-286 | 1.0014 | 1921-293 | 1.3100 | 0521-365 | 1.9680 |
| 0917+624 | 0.4472 | 1014+615 | 0.6437 | 1502+036 | 0.8044 | 1845+797 | 1.0022 | 1510-089 | 1.3121 | 1142+198 | 1.9859 |
| 0202+319 | 0.4481 | 1150+812 | 0.6446 | 2234+282 | 0.8048 | 1641+399 | 1.0024 | 2030+547 | 1.3128 | 1222+131 | 2.0200 |
| 0219+428 | 0.4486 | 1514+197 | 0.6497 | 0920+390 | 0.8095 | 1725+044 | 1.0128 | 1458+718 | 1.3160 | 0355+508 | 2.0327 |
| 1738+476 | 0.4550 | 0949+354 | 0.6540 | 1045-188 | 0.8108 | 1032-199 | 1.0129 | 0104-408 | 1.3182 | 1226+023 | 2.0502 |
| 1547+507 | 0.4623 | 1637+574 | 0.6544 | 0554+242 | 0.8109 | 0405-385 | 1.0158 | 2320-035 | 1.3197 | 1354-174 | 2.1224 |
| 0016+731 | 0.4627 | 0833+585 | 0.6573 | 0657+172 | 0.8133 | 2155-304 | 1.0167 | 1145-071 | 1.3201 | 0118-272 | 2.1310 |
| 1128+385 | 0.4635 | 1616+063 | 0.6597 | 1147+245 | 0.8157 | 0306+102 | 1.0201 | 1130+009 | 1.3270 | 0648-165 | 2.1341 |
| 1418+546 | 0.4642 | 0925-203 | 0.6627 | 1116+128 | 0.8201 | 0953+254 | 1.0235 | 1354-152 | 1.3290 | 1048-313 | 2.1452 |
| 1726+455 | 0.4652 | 0400+258 | 0.6647 | 1057-797 | 0.8218 | 0611+131 | 1.0266 | 1706-174 | 1.3326 | 0019+058 | 2.1814 |
| 1053+704 | 0.4654 | 0300+470 | 0.6649 | 1038+528 | 0.8266 | 1656+053 | 1.0406 | 1451-375 | 1.3481 | 0710+439 | 2.1996 |
| 0812+367 | 0.4657 | 0609+607 | 0.6696 | 1519-273 | 0.8283 | 2150+173 | 1.0458 | 1351-018 | 1.3682 | 0237-233 | 2.2393 |
| 0805+410 | 0.4689 | 2356+385 | 0.6722 | 1954-388 | 0.8288 | 1504-166 | 1.0502 | 2143-156 | 1.3717 | 1442+101 | 2.2633 |
| 0014+813 | 0.4704 | 0333+321 | 0.6740 | 0859+470 | 0.8306 | 0952+179 | 1.0516 | 1030+074 | 1.3763 | 2134+004 | 2.4165 |
| 1842+681 | 0.4802 | 1049+215 | 0.6771 | 1252+119 | 0.8326 | 0834-201 | 1.0522 | 1908-201 | 1.3788 | 1055+018 | 2.4609 |
| 0607-157 | 0.4804 | 0850+581 | 0.6798 | 1012+232 | 0.8389 | 1557+032 | 1.0611 | 0007+171 | 1.3895 | 0919-260 | 2.6582 |
| 1053+815 | 0.4890 | 1508-055 | 0.6802 | 0119+115 | 0.8408 | 2355-106 | 1.0655 | 0341+158 | 1.4014 | 1622-297 | 2.8362 |
| 0010+405 | 0.4901 | 1741-038 | 0.6823 | 0829+046 | 0.8413 | 0119-041 | 1.0674 | 1237-101 | 1.4073 | 2037-253 | 3.2076 |
| 2209+236 | 0.4904 | 1624+416 | 0.6829 | 1020+400 | 0.8445 | 0414-189 | 1.0679 | 2126-158 | 1.4093 | 0316+413 | 3.9531 |
| 0923+392 | 0.4912 | 1744+557 | 0.6859 | 2318+049 | 0.8483 | 1022+194 | 1.0704 | 1038+064 | 1.4095 | 2328+107 | 4.5366 |
| 1823+568 | 0.4914 | 2145+067 | 0.6882 | 0342+147 | 0.8491 | 2149+056 | 1.0739 | 1443-162 | 1.4209 | 0212+735 | 4.7541 |
| 2319+272 | 0.4947 | 1236+077 | 0.6907 | 0955+326 | 0.8502 | 1815-553 | 1.0751 | 0656+082 | 1.4403 | | |
| 0302+625 | 0.5029 | 0430+289 | 0.6909 | 1821+107 | 0.8510 | 0319+121 | 1.0775 | 1402+044 | 1.4439 | | |
| 0440+345 | 0.5030 | 0528+134 | 0.6911 | 1656+477 | 0.8569 | 2210-257 | 1.0833 | 1514-241 | 1.4470 | | |
| 0109+224 | 0.5079 | 0917+449 | 0.6916 | 1307+121 | 0.8592 | 0055+300 | 1.0854 | 2216-038 | 1.4487 | | |
| 2007+777 | 0.5098 | 0544+273 | 0.6958 | 1936-155 | 0.8592 | 1144-379 | 1.0862 | 1222+037 | 1.4584 | | |
| 0235+164 | 0.5122 | 2320+506 | 0.6972 | 2029+121 | 0.8607 | 0818-128 | 1.0894 | 0423+051 | 1.4636 | | |
| 0454+844 | 0.5123 | 0202+149 | 0.6986 | 1219+044 | 0.8608 | 0338-214 | 1.0896 | 0305+039 | 1.4690 | | |
| 0743+259 | 0.5194 | 0507+179 | 0.6999 | 0458-020 | 0.8627 | 0426-380 | 1.0900 | 1352-104 | 1.4696 | | |
| 1324+224 | 0.5257 | 0735+178 | 0.7014 | 0745+241 | 0.8643 | 0422+004 | 1.0923 | 0150-334 | 1.4886 | | |
| 1150+497 | 0.5321 | 0159+723 | 0.7073 | 1427+543 | 0.8670 | 1347+539 | 1.0955 | 0821+394 | 1.4931 | | |
| 0420+417 | 0.5366 | 1308+328 | 0.7086 | 0146+056 | 0.8713 | 0742+103 | 1.1185 | 2329-162 | 1.4937 | | |
| 0446+112 | 0.5417 | 1101+384 | 0.7113 | 0248+430 | 0.8725 | 0906+015 | 1.1186 | 0605-085 | 1.4956 | | |
| 1749+096 | 0.5418 | 2200+420 | 0.7115 | 0208-512 | 0.8779 | 0434-188 | 1.1187 | 1243-072 | 1.4963 | | |
| 1606+106 | 0.5429 | 1216+487 | 0.7124 | 1334-127 | 0.8789 | 2302+232 | 1.1314 | 1034-293 | 1.5029 | | |
| 0718+792 | 0.5430 | 1738+499 | 0.7129 | 0046+316 | 0.8789 | 1406-076 | 1.1369 | 1428+422 | 1.5244 | | |
| 0707+476 | 0.5434 | 1610-771 | 0.7169 | 2244-372 | 0.8815 | 1257+145 | 1.1383 | 1730-130 | 1.5315 | | |
| 1030+415 | 0.5456 | 0808+019 | 0.7192 | 1424-418 | 0.8834 | 1656+348 | 1.1414 | 0106+013 | 1.5450 | | |
| 1156+295 | 0.5468 | 2214+350 | 0.7195 | 0229+131 | 0.8860 | 1244-255 | 1.1501 | 1647-296 | 1.5457 | | |
| 1308+326 | 0.5500 | 0537-441 | 0.7204 | 1901+319 | 0.8898 | 1622-253 | 1.1504 | 1502+106 | 1.5470 | | |
| 1745+624 | 0.5524 | 0827+243 | 0.7251 | 0309+411 | 0.8943 | 1219+285 | 1.1524 | 1937-101 | 1.5495 | | |
| 1642+690 | 0.5548 | 1856+737 | 0.7296 | 1240+381 | 0.8962 | 0620+389 | 1.1544 | 0317+188 | 1.5538 | | |

Table 4.  Ranking of sources by WRMS+WADEV, all sessions 652 sources.

| Source | Value | Source | Value | Source | Value | Source | Value | Source | Value | Source | Value |
|---|---|---|---|---|---|---|---|---|---|---|---|
| 1330+476 | 0.1961 | 2007+777 | 0.6117 | 2358+189 | 0.8194 | 0201+113 | 1.1122 | 0656+082 | 1.4235 | 0521-365 | 1.8970 |
| 1357+769 | 0.2860 | 1022+194 | 0.6118 | 0119+115 | 0.8228 | 0341+158 | 1.1135 | 0642+214 | 1.4279 | 0859-140 | 1.9014 |
| 0836+182 | 0.2924 | 1150+812 | 0.6119 | 1413+135 | 0.8250 | 1213-172 | 1.1140 | 0111+021 | 1.4341 | 0415+379 | 1.9085 |
| 0834+250 | 0.3118 | 1954+513 | 0.6129 | 0534-611 | 0.8274 | 0522-611 | 1.1170 | 1611-710 | 1.4344 | 0524-460 | 1.9129 |
| 1300+580 | 0.3126 | 0234+285 | 0.6140 | 0605-085 | 0.8286 | 0539-057 | 1.1176 | 0405-385 | 1.4357 | 0637-337 | 1.9145 |
| 0059+581 | 0.3202 | 0812+367 | 0.6144 | 0601+245 | 0.8321 | 2149-307 | 1.1178 | 1647-296 | 1.4380 | 2233-148 | 1.9159 |
| 0133+476 | 0.3235 | 0748+126 | 0.6165 | 1656+477 | 0.8346 | 0506-612 | 1.1193 | 1510-089 | 1.4382 | 0831+557 | 1.9287 |
| 1807+698 | 0.3303 | 0821+394 | 0.6166 | 1806+456 | 0.8358 | 2230+114 | 1.1247 | 0116-219 | 1.4400 | 1239+376 | 1.9335 |
| 0602+673 | 0.3403 | 1754+155 | 0.6168 | 2253+417 | 0.8380 | 0537-286 | 1.1262 | 1514-241 | 1.4536 | 0919-260 | 1.9431 |
| 0759+183 | 0.3412 | 1342+662 | 0.6196 | 2234+282 | 0.8388 | 1101-536 | 1.1265 | 2143-156 | 1.4547 | 1104-445 | 1.9513 |
| 0515+208 | 0.3450 | 2229+695 | 0.6203 | 1821+107 | 0.8429 | 0130-171 | 1.1342 | 1111+149 | 1.4549 | 0425+048 | 1.9519 |
| 0749+540 | 0.3487 | 0820+560 | 0.6249 | 0221+067 | 0.8455 | 1251-713 | 1.1349 | 0241+622 | 1.4553 | 1048-313 | 1.9568 |
| 0642+449 | 0.3578 | 0302+625 | 0.6260 | 1936-155 | 0.8534 | 1219+285 | 1.1350 | 2109-811 | 1.4617 | 0534-340 | 1.9612 |
| 0552+398 | 0.3628 | 0536+145 | 0.6280 | 1042+071 | 0.8538 | 2335-027 | 1.1382 | 0202-172 | 1.4626 | 0423+233 | 1.9904 |
| 1739+522 | 0.3889 | 1538+149 | 0.6297 | 1901+319 | 0.8545 | 1020+400 | 1.1436 | 0131-522 | 1.4708 | 1222+131 | 1.9906 |
| 0955+476 | 0.3922 | 2150+173 | 0.6337 | 0426+273 | 0.8567 | 0248+430 | 1.1513 | 1038+064 | 1.4713 | 1328+307 | 2.0167 |
| 1642+690 | 0.3947 | 1749+701 | 0.6338 | 1124-186 | 0.8612 | 0235-618 | 1.1521 | 1937-101 | 1.4738 | 1519-294 | 2.0229 |
| 0110+495 | 0.4101 | 0707+476 | 0.6345 | 0745-241 | 0.8619 | 2131-021 | 1.1521 | 0338-214 | 1.4743 | 1420-679 | 2.0404 |
| 1638+398 | 0.4113 | 1508+572 | 0.6364 | 1519-273 | 0.8661 | 0814+425 | 1.1522 | 0104-408 | 1.4745 | 2255-282 | 2.0450 |
| 1011+250 | 0.4151 | 1624+416 | 0.6373 | 2318+049 | 0.8668 | 0834-201 | 1.1545 | 1435-218 | 1.4816 | 2037-253 | 2.0720 |
| 1044+719 | 0.4222 | 1502+036 | 0.6382 | 1427+543 | 0.8670 | 0039+230 | 1.1561 | 0047-579 | 1.4836 | 0826-373 | 2.0746 |
| 1417+385 | 0.4305 | 1717+178 | 0.6403 | 1213+350 | 0.8686 | 1511-100 | 1.1575 | 2216-038 | 1.4902 | 1433+304 | 2.0771 |
| 0420+417 | 0.4326 | 0400+258 | 0.6412 | 1219+044 | 0.8693 | 0422+004 | 1.1583 | 2321-375 | 1.4911 | 0332-403 | 2.0866 |
| 1803+784 | 0.4329 | 1216+487 | 0.6445 | 1123+264 | 0.8742 | 0406-121 | 1.1592 | 1502-106 | 1.4937 | 1012-448 | 2.0953 |
| 0804+499 | 0.4339 | 0430+289 | 0.6447 | 2300-683 | 0.8771 | 1130+009 | 1.1602 | 0710+439 | 1.5063 | 2331-240 | 2.1298 |
| 1611+343 | 0.4377 | 0212+735 | 0.6461 | 0309+411 | 0.8774 | 0138-097 | 1.1651 | 0220-349 | 1.5091 | 1929-457 | 2.1339 |
| 1324+224 | 0.4443 | 0322+222 | 0.6464 | 0730-504 | 0.8795 | 1555+001 | 1.1653 | 1556-245 | 1.5101 | 1354-174 | 2.1408 |
| 0446+112 | 0.4464 | 1101+384 | 0.6478 | 0637-752 | 0.8834 | 0007+171 | 1.1655 | 1127-145 | 1.5153 | 2355-534 | 2.1579 |
| 1636+473 | 0.4489 | 1150+497 | 0.6513 | 1038+528 | 0.8843 | 2155-152 | 1.1662 | 1424-366 | 1.5154 | 1143-287 | 2.1959 |
| 0538+498 | 0.4499 | 1147+245 | 0.6527 | 0951+693 | 0.8845 | 0305+039 | 1.1687 | 0648-165 | 1.5167 | 0511-220 | 2.1985 |
| 0917+624 | 0.4504 | 1923+210 | 0.6553 | 0414-189 | 0.8860 | 1824-582 | 1.1691 | 2048+312 | 1.5238 | 0056-001 | 2.2048 |
| 1128+385 | 0.4507 | 0333+321 | 0.6570 | 1228+126 | 0.8862 | 2023+335 | 1.1727 | 1416+067 | 1.5256 | 1511-476 | 2.2184 |
| 2037+511 | 0.4574 | 2250+190 | 0.6582 | 0610+260 | 0.8936 | 0506+101 | 1.1852 | 2126-158 | 1.5298 | 2054-377 | 2.2378 |
| 1617+229 | 0.4588 | 2200+420 | 0.6603 | 0229+131 | 0.8950 | 1614+051 | 1.1856 | 1155-251 | 1.5299 | 0529+075 | 2.2391 |
| 1842+681 | 0.4685 | 1744+557 | 0.6637 | 1633+382 | 0.8959 | 0440-003 | 1.1896 | 0438-436 | 1.5300 | 1038+529 | 2.2747 |
| 1418+546 | 0.4744 | 0003+380 | 0.6686 | 1226+373 | 0.8999 | 1954-388 | 1.1907 | 1622-297 | 1.5331 | 1055-301 | 2.2837 |
| 0805+410 | 0.4770 | 2351-154 | 0.6686 | 1049+215 | 0.9003 | 2355-106 | 1.1922 | 1445-161 | 1.5342 | 1253-055 | 2.2930 |
| 0440+345 | 0.4831 | 0549-575 | 0.6705 | 0458-020 | 0.9029 | 0530-727 | 1.1932 | 1815-553 | 1.5373 | 1245-457 | 2.2974 |
| 0035+413 | 0.4847 | 0736+017 | 0.6757 | 0945+408 | 0.9069 | 0723-008 | 1.1955 | 2052-474 | 1.5385 | 0743-673 | 2.3320 |
| 0358+210 | 0.4870 | 0123+257 | 0.6759 | 0611+131 | 0.9100 | 0823-223 | 1.1956 | 1257-145 | 1.5418 | 0108-388 | 2.3399 |
| 0010+405 | 0.4900 | 0912+029 | 0.6762 | 1420+326 | 0.9119 | 1557+032 | 1.1967 | 0402-362 | 1.5459 | 1142+198 | 2.3659 |
| 2209+236 | 0.4901 | 2201+315 | 0.6766 | 1404+286 | 0.9136 | 2353-686 | 1.2009 | 1133+704 | 1.5649 | 2058-297 | 2.3859 |
| 1751+288 | 0.4927 | 1417+273 | 0.6774 | 1932-204 | 0.9147 | 1428+422 | 1.2024 | 1921-293 | 1.5666 | 1451-400 | 2.4187 |
| 1800+440 | 0.4941 | 1514+197 | 0.6822 | 1546+027 | 0.9156 | 2005+642 | 1.2024 | 1104+728 | 1.5785 | 1016-311 | 2.4262 |
| 0805-077 | 0.4948 | 1435+638 | 0.6889 | 1624-617 | 0.9186 | 0430+052 | 1.2072 | 0400-319 | 1.5876 | 0002-478 | 2.4288 |
| 1823+568 | 0.4959 | 0808+019 | 0.6902 | 1705+135 | 0.9257 | 0208-512 | 1.2089 | 2252-090 | 1.5997 | 2227-088 | 2.4628 |
| 1726+455 | 0.4969 | 0821+621 | 0.6926 | 1423+146 | 0.9327 | 2149+056 | 1.2117 | 0153+744 | 1.6027 | 1349-439 | 2.4712 |
| 0235+164 | 0.5013 | 0716+714 | 0.6930 | 0823+033 | 0.9327 | 1237-101 | 1.2120 | 1128-047 | 1.6077 | 1148-671 | 2.5248 |
| 0219+428 | 0.5040 | 0403-132 | 0.7004 | 0454-844 | 0.9381 | 1548+056 | 1.2153 | 1817-254 | 1.6104 | 2314-340 | 2.5319 |
| 0014+813 | 0.5075 | 0954+658 | 0.7006 | 1252+119 | 0.9399 | 0405-123 | 1.2156 | 0451-282 | 1.6117 | 0711+356 | 2.5401 |
| 1432+200 | 0.5128 | 1308+328 | 0.7045 | 0952+179 | 0.9450 | 2008-159 | 1.2176 | 1034-293 | 1.6140 | 2254+024 | 2.5415 |
| 0743+259 | 0.5189 | 2356+385 | 0.7071 | 1338+381 | 0.9453 | 0319+121 | 1.2190 | 1040+244 | 1.6160 | 1412-368 | 2.5465 |
| 0151+474 | 0.5197 | 0528+134 | 0.7081 | 2021+317 | 0.9457 | 0906+015 | 1.2201 | 2128-123 | 1.6269 | 2030-689 | 2.5678 |
| 0202+319 | 0.5206 | 0107-610 | 0.7108 | 2254+074 | 0.9472 | 1758-651 | 1.2248 | 1004-500 | 1.6274 | 1226-028 | 2.6455 |
| 0016+731 | 0.5206 | 0850+581 | 0.7113 | 1045-188 | 0.9489 | 0256+075 | 1.2257 | 2251+158 | 1.6307 | 1725-795 | 2.6640 |
| 0607-157 | 0.5208 | 0547+234 | 0.7121 | 2244-372 | 0.9523 | 2204-540 | 1.2300 | 0454-463 | 1.6314 | 1633-810 | 2.6830 |
| 0923+392 | 0.5209 | 2029+121 | 0.7128 | 0239+108 | 0.9525 | 0112-017 | 1.2302 | 0150-334 | 1.6370 | 1226+023 | 2.6926 |
| 0738+491 | 0.5245 | 1144+402 | 0.7197 | 2235+731 | 0.9527 | 0013-005 | 1.2325 | 1302-102 | 1.6395 | 0912+297 | 2.7101 |
| 2000+472 | 0.5254 | 1746+470 | 0.7199 | 0743-006 | 0.9531 | 1032-199 | 1.2386 | 2000-330 | 1.6476 | 2059-786 | 2.7364 |
| 0836+710 | 0.5336 | 0215+015 | 0.7239 | 0048-097 | 0.9550 | 1831-711 | 1.2419 | 1951+355 | 1.6536 | 1806-458 | 2.7744 |
| 1014+615 | 0.5338 | 1741-038 | 0.7241 | 1958-179 | 0.9566 | 2243-123 | 1.2425 | 0048-427 | 1.6597 | 0122-514 | 2.7879 |
| 2320+506 | 0.5357 | 0600+177 | 0.7253 | 0055+300 | 0.9583 | 1443-162 | 1.2435 | 0238-084 | 1.6597 | 2134+004 | 2.7893 |
| 1012+232 | 0.5407 | 2017+745 | 0.7255 | 1656+053 | 0.9600 | 2328+107 | 1.2455 | 2002-375 | 1.6630 | 0629+160 | 2.7935 |
| 0159+723 | 0.5412 | 0202+149 | 0.7256 | 1334-127 | 0.9644 | 0423+051 | 1.2490 | 1619-680 | 1.6634 | 1021-006 | 2.8242 |
| 1039+811 | 0.5444 | 1637+826 | 0.7277 | 0234-301 | 0.9714 | 2059+034 | 1.2526 | 2312-319 | 1.6747 | 0920-397 | 2.8557 |
| 1749+096 | 0.5453 | 0828+493 | 0.7283 | 1903-802 | 0.9764 | 1727+502 | 1.2533 | 1730-130 | 1.6753 | 1451-375 | 2.8818 |
| 1600+335 | 0.5478 | 0615+820 | 0.7292 | 0007+106 | 0.9814 | 0252-549 | 1.2535 | 1325-558 | 1.6790 | 2005+403 | 2.9384 |
| 1639+230 | 0.5490 | 2144+092 | 0.7337 | 1022-665 | 0.9862 | 2344+092 | 1.2552 | 2250+194 | 1.7003 | 0629-418 | 2.9566 |
| 2236-572 | 0.5494 | 0827+243 | 0.7365 | 0620+389 | 0.9873 | 1004+141 | 1.2626 | 1222+037 | 1.7109 | 2306-312 | 2.9955 |
| 1236+077 | 0.5507 | 1240+381 | 0.7373 | 0459+135 | 0.9894 | 1657-562 | 1.2674 | 2142-758 | 1.7109 | 1604-333 | 3.0199 |
| 0949+354 | 0.5508 | 0754+100 | 0.7411 | 1354-152 | 0.9897 | 2325-150 | 1.2675 | 0459+060 | 1.7140 | 0118-272 | 3.0401 |
| 0149+218 | 0.5517 | 0444+634 | 0.7417 | 1315+346 | 0.9928 | 2021+614 | 1.2679 | 1657-261 | 1.7250 | 0733-174 | 3.0773 |
| 1053+815 | 0.5533 | 0300+470 | 0.7434 | 0306+102 | 0.9957 | 0334-546 | 1.2696 | 1402-012 | 1.7286 | 0437-454 | 3.0991 |
| 1738+476 | 0.5550 | 1406-076 | 0.7436 | 1409+218 | 0.9960 | 2008-068 | 1.2707 | 1925-610 | 1.7545 | 2102-659 | 3.1410 |
| 1547+507 | 0.5551 | 1347+539 | 0.7444 | 1656+348 | 1.0017 | 2320-035 | 1.2747 | 0537-158 | 1.7590 | 2146-783 | 3.2624 |
| 1606+106 | 0.5569 | 1342+663 | 0.7467 | 0003-066 | 1.0054 | 0738-674 | 1.2973 | 1101-325 | 1.7616 | 0518+165 | 3.3091 |

| | | | | | | | | | |
|---|---|---|---|---|---|---|---|---|---|
| 1734+363 | 0.5587 | 0502+049 | 0.7468 | 0454-810 | 1.0076 | 1156-663 | 1.3020 | 0646-306 | 1.7696 | 2352+495 | 3.3987 |
| 0406-127 | 0.5595 | 1504+377 | 0.7503 | 2121+053 | 1.0101 | 1402+044 | 1.3025 | 2337+264 | 1.7698 | 0405+304 | 3.4012 |
| 1745+624 | 0.5604 | 1637+574 | 0.7516 | 0554+242 | 1.0130 | 1641+399 | 1.3027 | 2106-413 | 1.7728 | 1345+125 | 3.4489 |
| 1030+415 | 0.5606 | 1929+226 | 0.7525 | 1034-374 | 1.0131 | 0008-264 | 1.3040 | 0528-250 | 1.7775 | 1105-680 | 3.4543 |
| 1459+480 | 0.5606 | 0650+371 | 0.7546 | 0230-790 | 1.0167 | 0317+188 | 1.3085 | 2210-257 | 1.7799 | 0335-364 | 3.5931 |
| 1849+670 | 0.5630 | 1057-797 | 0.7595 | 1830+285 | 1.0174 | 1023+131 | 1.3098 | 1313-333 | 1.7817 | 1607+268 | 3.6208 |
| 0224+671 | 0.5641 | 0920+390 | 0.7634 | 0537-441 | 1.0182 | 1458+718 | 1.3117 | 0738+313 | 1.7824 | 1116-462 | 3.6519 |
| 1508-055 | 0.5669 | 0458+138 | 0.7644 | 0529+483 | 1.0196 | 1622-253 | 1.3218 | 0326+277 | 1.7862 | 1826+796 | 3.7077 |
| 1221+809 | 0.5695 | 0336-019 | 0.7645 | 0953+254 | 1.0230 | 1040+123 | 1.3260 | 2344-514 | 1.7862 | 0503-608 | 3.7218 |
| 1448+762 | 0.5733 | 1751+441 | 0.7652 | 1610-771 | 1.0295 | 0237-027 | 1.3272 | 1148-001 | 1.7875 | 1156-094 | 3.8763 |
| 2319+272 | 0.5736 | 0657+172 | 0.7659 | 0454-234 | 1.0305 | 1030+074 | 1.3302 | 1255-316 | 1.8046 | 1454-354 | 3.9293 |
| 1053+704 | 0.5756 | 2145+067 | 0.7661 | 0516-621 | 1.0342 | 1354+195 | 1.3326 | 1706-174 | 1.8076 | 0237-233 | 4.0042 |
| 2136+141 | 0.5774 | 1856+737 | 0.7687 | 1145-071 | 1.0408 | 1424-418 | 1.3336 | 2155-304 | 1.8175 | 0026+346 | 4.0215 |
| 0833+585 | 0.5789 | 0507+179 | 0.7693 | 1926+087 | 1.0431 | 0818-128 | 1.3344 | 0355+508 | 1.8220 | 1236-684 | 4.0650 |
| 1308+326 | 0.5798 | 1215+303 | 0.7721 | 0148+274 | 1.0460 | 2329-162 | 1.3354 | 1933-400 | 1.8319 | 0236+610 | 4.3754 |
| 2113+293 | 0.5798 | 0046+316 | 0.7727 | 0135-247 | 1.0461 | 1243-072 | 1.3419 | 0457+024 | 1.8340 | 0316+413 | 4.5249 |
| 1156+295 | 0.5816 | 1532+016 | 0.7755 | 0237+040 | 1.0515 | 1549-790 | 1.3469 | 0113-118 | 1.8376 | 1043+066 | 4.6841 |
| 2201+171 | 0.5823 | 1116+128 | 0.7790 | 0636+680 | 1.0599 | 2051+745 | 1.3470 | 1522+155 | 1.8385 | 1718-649 | 4.8314 |
| 0109+224 | 0.5846 | 0735+178 | 0.7816 | 1448-648 | 1.0606 | 1351-018 | 1.3478 | 2329-384 | 1.8416 | 2128+048 | 4.8494 |
| 0609+607 | 0.5846 | 0859+470 | 0.7850 | 0119+041 | 1.0700 | 1015+359 | 1.3520 | 2245-328 | 1.8429 | 1947+079 | 4.9357 |
| 1705+456 | 0.5865 | 1652+398 | 0.7865 | 0839+187 | 1.0722 | 1144-379 | 1.3573 | 2351-309 | 1.8430 | 1031+567 | 5.2445 |
| 1758+388 | 0.5867 | 0805+046 | 0.7865 | 1659-621 | 1.0748 | 0422-380 | 1.3747 | 1430+365 | 1.8434 | 0134+329 | 6.0889 |
| 1655+077 | 0.5884 | 0544+273 | 0.7897 | 2030+547 | 1.0750 | 0037+139 | 1.3783 | 2106+143 | 1.8435 | 1322-427 | 7.8807 |
| 1732+389 | 0.5902 | 0727-115 | 0.7903 | 1307+121 | 1.0788 | 1424+240 | 1.3783 | 1143-696 | 1.8472 | 2005-489 | 8.1783 |
| 2351+456 | 0.5927 | 0829+046 | 0.7904 | 1725+044 | 1.0793 | 1129-580 | 1.3818 | 0500+019 | 1.8581 | 1323+321 | 12.8057 |
| 0718+792 | 0.5929 | 0409+229 | 0.7912 | 0308-611 | 1.0803 | 1143-245 | 1.3819 | 0355-669 | 1.8701 | 0218+357 | 47.0606 |
| 0722+145 | 0.5960 | 0556+238 | 0.7935 | 1705+018 | 1.0874 | 1352-104 | 1.3837 | 2326-477 | 1.8721 | | |
| 0917+449 | 0.6005 | 1616+063 | 0.7941 | 1244-255 | 1.0939 | 0420-014 | 1.3839 | 0434-188 | 1.8726 | | |
| 2223-052 | 0.6009 | 1928+738 | 0.8027 | 1055+018 | 1.0975 | 1504-166 | 1.4003 | 0426-380 | 1.8842 | | |
| 0508+138 | 0.6046 | 0651+410 | 0.8031 | 1920-211 | 1.0999 | 1442+101 | 1.4084 | 2227-399 | 1.8867 | | |
| 2214+350 | 0.6050 | 1743+173 | 0.8073 | 2302+232 | 1.1074 | 1908-201 | 1.4144 | 2232-488 | 1.8869 | | |
| 0851+202 | 0.6085 | 0602+405 | 0.8106 | 0106+013 | 1.1090 | 0019+058 | 1.4154 | 0524-485 | 1.8902 | | |
| 0955+326 | 0.6087 | 0146+056 | 0.8141 | 0742+103 | 1.1100 | 1845+797 | 1.4156 | 1935-692 | 1.8924 | | |
| 1738+499 | 0.6115 | 0342+147 | 0.8145 | 0302-623 | 1.1120 | 0925-203 | 1.4167 | 2345-167 | 1.8955 | | |

## 4. Sources with non-random position variations

MacMillan & Ma (2007) identified eight sources with irregular apparent motion: 0014+813, 0528+134, 0923+392 (4C39.25), 1739+522, 2145+067, 2223–052 (3C446), 2234+282 and 2243–123. In addition, visual examination of time series gsf003a, usn000d and usn001a (the latter was used for control only; as mentioned above, this series along with a couple of others show many change points in position time series) allows us to select the following sources for further investigation (as a rule, only change points after 1990.0 were considered): 0003-066, 0007+171, 0106+013, 0119+015, 0133+476, 0149+218, 0201+113, 0202+149, 0235+164, 0336–019, 0420–014, 0451–282, 0454–234, 0556+238, 0602+673, 0953+254, 1044+719, 1308+326, 1610–771, 1611+343, 1741–038, 2121+053, 2128–123, 2136+141, 2201+315.